\documentclass{aa}  
\usepackage{graphicx}
\usepackage{txfonts}
\begin{document}
   \title{Periodic radial velocity variations in RU Lupi
          \thanks{Based
	  on observations collected at the European Southern Observatory, Chile
	  (proposals 65.I-0404, 69.C-0481 and 75.C-0292(A))}
          }

%  \subtitle{}

   \author{H. C. Stempels\inst{1}
           \and
	   G. F. Gahm\inst{2}
	   \and
	   P. P. Petrov\inst{3}
	   }

   \offprints{H. C. Stempels}

   \institute{School of Physics \& Astronomy, University of St Andrews, North
              Haugh, St Andrews KY16 9SS, Scotland,\\
              email: \mbox{Eric.Stempels@st-andrews.ac.uk}
              \and
	      Stockholm Observatory, AlbaNova University Centre,
	      SE-106 91 Stockholm, Sweden
	      \and
	      Crimean Astrophysical Observatory, p/o Nauchny, Crimea, 98409
	      Ukraine
	      }

   \date{Accepted for publication by A\&A}

   % \abstract{}{}{}{}{} 
   % 5 {} token are mandatory

   \abstract
   {
   RU Lup is a Classical T Tauri star with unusually strong emission
   lines, which has been interpreted as manifestations of accretion. Recently,
   evidence has accumulated that this star might have a variable radial
   velocity.
   }
   {We intended to investigate in more detail the possible variability in
   radial velocity using a set of 68 high-resolution spectra taken at the VLT
   (UVES), the AAT (UCLES) and the CTIO (echelle).
   }
   {Using standard cross-correlation techniques, we determined the radial
   velocity of RU~Lup. We analysed these results with Phase-dispersion
   minimization and the Lomb-Scargle periodogram and searched for possible
   periodicities in the obtained radial velocities. We also analysed
   changes in the absorption line shapes and the photometric variability of RU
   Lup.}
   {Our analysis indicated that RU Lup exhibits variations in radial
   velocity with a periodicity of 3.71 days and an amplitude of $2.17$
   km~s${}^{-1}$. These variations can be explained by the presence of large
   spots, or groups of spots, on the surface of RU Lup. We also considered a
   low-mass companion and stellar pulsations as alternative sources for these
   variations but found these to be unlikely.}
   


   \maketitle
%
%________________________________________________________________

\section{Introduction}

T Tauri stars (TTS) show hydrogen emission lines superimposed on relatively
normal late-type photospheric spectra in the optical spectral region. A fraction
of the TTS have very strong and broad emission lines, and in addition emission
lines of neutral and once ionised metals, and in some cases He. These stars are
nowadays referred to as the classical T~Tauri stars (CTTS) since they were the
first to be studied spectroscopically by Joy (\cite{joy}). CTTS are identified
with pre-main-sequence low-mass stars, and many of their observed
characteristics can be explained within the framework of magnetospheric
accretion (see for example, Hartmann \cite{hartmann}). According to this model,
gas is ionised at the interior of a circumstellar, dusty disk, after which the
plasma is accreted along stellar dipole magnetic field lines, and eventually
dumped under virtually free fall at high latitudes, creating a shocked region
that heats the photosphere locally. The energy released in this hot accretion
shock then manifests itself as excess continuous emission.
The model of magnetospheric accretion has gained support through the detection
of relatively strong (several kG) magnetic fields around CTTS (Guenther et al.
\cite{guenther99}; Johns-Krull et al. \cite{johns99a}; Valenti \& Johns-Krull
\cite{valenti04}).

Among the CTTS, there is a small group of stars with extremely rich emission
line spectra and veiling, which on occasions can obliterate the photospheric
spectrum completely. Some of these ``enhanced'' CTTS vary dramatically and
randomly in brightness, line profiles and veiling on time scales as short as
hours, which is often interpreted as being related to changes in the
instantaneous accretion rate. Still, what the driving force is behind these
strongly enhanced accretion signatures is unclear.

Several of the enhanced CTTS have turned out to be close, double-lined
spectroscopic binaries. In DQ Tau (Basri et al. \cite {basri}), UZ Tau E (Martin
et al. \cite{martin}) and V4046 Sgr (Stempels \& Gahm \cite{st04}), accretion
signatures appear to be modulated by the binary orbit. Other ``enhanced'' CTTS
show velocity changes of small amplitude in single photospheric lines and may be
single-lined spectroscopic binaries (Melo \cite{me03}; Gahm \cite{gahm06}). It
is also possible that the changes in radial velocity are caused by starspots
moving across the stellar disk. It may be difficult to discriminate between the
two scenarios. An example is the extreme CTTS RW Aur A for which Petrov et al.
(\cite{petrov01}) considered a low-mass binary companion. However, they also
found reasonable fits to the line variations by modeling spot-induced absorption
line variations.

Within this framework we will discuss the well-known and extremely active CTTS
RU Lup, for which we recently discovered a possible periodicity in radial
velocity (Gahm et al. \cite{gahm05}). RU Lup is well-studied both
spectroscopically and photometrically, displaying very strong and variable
accretion signatures. The properties of RU Lup were described in depth by
Herczeg et al. (\cite{herczeg05}) and Stempels \& Piskunov (\cite{st02}).

\section{Observations}

We observed RU Lup with the UVES spectrograph at the \mbox{8-m} VLT/UT2 in Chile
during three observing runs, April 15\,--\,16, 2000, April 15\,--\,18, 2002 and 
August 14\,--\,17, 2005. We used UVES in dichroic-mode, to obtain simultaneous
exposures with the blue and red arm of the spectrograph. In this way, the
wavelength coverage is almost continuous, and ranges from $3500$ to $6700$
{\AA}. High-quality spectra, normally at a resolution of $R \approx 60\,000$,
could be obtained in exposures as short as 15\,--\,30 minutes with the blue arm
(S/N-ratio $\sim 50$), and $10$ minutes with the red arm (S/N-ratio $\sim 150$).
In total we obtained 29 blue and 66 red spectra.

In addition to the UVES data, we also obtained one spectrum of RU Lup taken
with the echelle spectrograph at the \mbox{4-m} CTIO, on July 17, 1998,
and one spectrum taken with the UCLES spectrograph at the \mbox{4-m} AAT, on July
11, 2005.

All data were reduced with the IDL-based package REDUCE (Piskunov \& Valenti
\cite{pi02}). This package uses advanced techniques for accurate
bias-subtraction, modelling of scattered light and flat-fielding, and performs
optimal order extraction through 2D modelling of the individual orders.

\section{Analysis}

\subsection{A synthetic template spectrum for RU Lup}

The spectrum of RU Lup contains a large number of strong emission lines of e.g.
H, Ca {\sc ii} H \& K, Fe {\sc i} and {\sc ii}, [O {\sc i}] and [S~{\sc ii}]. 
Although line emission and veiling contaminate large parts of the spectrum,
there are shorter regions where narrow, photospheric absorption lines can be
seen. In these regions we determined the photospheric contribution in RU Lup by
calculating synthetic template spectra (see Stempels \& Piskunov \cite{st02},
\cite{st03}). This method assumes that the ``stellar'' spectrum can be described
by the sum of a featureless excess continuum and a normal late-type photospheric
spectrum. The model atmosphere, stellar parameters and atomic line data used are
the same as in Stempels \& Piskunov (\cite{st03}). Other advantages of a
synthetic template spectrum is that it allows to accurately determine the
stellar radial velocity and absorption line shapes from cross-correlation.

\begin{table}
\caption{Modified Julian Date, measured radial velocities and the Bisector
Inverse Slope (BIS, see Sect. \ref{sec:bisectors}) for RU~Lup. The error
($\sigma$) of our radial velocity measurements is $0.16$~km~s${}^{-1}$. Values
marked with ${}^*$ are taken from Melo (\cite{me03}), but transferred to MJD.
The observation marked with ${}^{**}$ was obtained at the CTIO, and the
observations marked with ${}^{***}$ at the AAT.}
\label{tab:radvel}
\centering 
\begin{tabular}{l r r | l r r}
\hline\hline
MJD & $v_{\rm rad}$   & BIS & MJD & $v_{\rm rad}$   & BIS\\
    & (km s${}^{-1}$) &     &     & (km s${}^{-1}$) & \\
\hline
51003.2106 & -1.92${}^*$    &	&   52381.215 &  -2.58  &   0.86  \\   
51316.1549 & -1.77${}^*$    &	&   52381.301 &  -2.38  &   1.11  \\   
51321.1779 & -0.35${}^*$    &	&   52381.309 &  -2.36  &   1.43  \\   
51666.2141 & -1.02${}^*$    &	&   52381.312 &  -2.35  &   0.80  \\   
51672.1329 & -2.48${}^*$    &	&   52381.383 &  -2.40  &   0.85  \\   
51674.2709 &  0.49${}^*$    &	&   52381.391 &  -2.28  &   0.51  \\   
51685.1903 &  0.47${}^*$    &	&   52381.398 &  -2.35  &   0.42  \\   
          &		    &	&   52382.203 &  -0.57  &  -1.12  \\   
51651.211 &  -0.38  &	0.12	&   52382.211 &  -0.25  &  -1.52  \\	      
51651.219 &  -0.46  &	0.04	&   52382.219 &  -0.37  &  -1.11  \\	      
51651.227 &  -0.60  &	0.33	&   52382.293 &  -0.19  &  -1.00  \\	      
51651.262 &  -0.80  &  -0.01	&   52382.301 &  -0.18  &  -1.05  \\	      
51651.270 &  -0.73  &  -0.09	&   52382.309 &  -0.19  &  -1.19  \\	      
51651.277 &  -0.67  &  -0.02	&   52382.359 &  -0.19  &  -0.90  \\	      
51651.316 &  -0.48  &	0.02	&   52382.367 &  -0.23  &  -0.78  \\	      
51651.324 &  -0.43  &	0.38	&   52382.375 &  -0.22  &  -0.97  \\	      
51651.328 &  -0.34  &	0.16	&   52383.203 &   0.16  &  -0.52  \\	      
51651.414 &  -0.33  &  -0.11	&   52383.211 &   0.06  &  -0.61  \\	      
51651.422 &  -0.35  &	0.52	&   52383.219 &   0.41  &  -0.91  \\	      
51652.184 &   1.04  &  -0.50	&   52383.301 &  -0.50  &  -0.79  \\	      
51652.191 &   1.08  &  -0.26	&   52383.309 &  -0.33  &  -0.68  \\	      
51652.195 &   0.96  &  -0.28	&   52383.316 &  -0.52  &  -0.62  \\	      
51652.293 &   0.67  &  -0.59	&   52383.367 &  -0.50  &  -0.89  \\	      
51652.301 &   0.65  &  -0.33	&   52383.375 &  -0.46  &  -0.94  \\	      
51652.309 &   0.60  &  -0.12	&   52383.383 &  -0.56  &  -0.83  \\	      
51652.371 &   0.43  &  -0.42	&   53596.992 &  -2.19  &   0.93  \\	      
51652.379 &   0.13  &  -0.01	&   53597.051 &  -2.90  &  -0.17  \\	      
51652.387 &   0.50  &  -0.15	&   53598.039 &  -2.74  &   0.74  \\	      
52380.207 &  -3.38  &  -0.11	&   53598.082 &  -2.62  &   0.57  \\	      
52380.215 &  -3.52  &	0.11	&   53598.969 &  -0.13  &  -0.52  \\	      
52380.223 &  -3.59  &	0.29	&   53599.016 &   0.16  &  -0.89  \\	      
52380.297 &  -3.74  &	1.04	&   53599.066 &   0.40  &  -0.59  \\	      
52380.305 &  -3.79  &	1.16	&   53599.969 &   1.06  &  -0.38  \\	      
52380.312 &  -3.59  &	0.41	&   53600.000 &   1.14  &  -0.94  \\	      
52380.383 &  -4.03  &	0.96	&   53600.062 &   0.42  &  -0.48  \\	      
52380.391 &  -3.87  &	0.90	&	      &		  \\		   
52380.398 &  -3.94  &	1.52	&   51011.984 &  -2.55${}^{**}$  \\	   
52381.199 &  -2.42  &	1.12	&	      &		  \\		   
52381.207 &  -2.57  &	0.79	&   53562.373 &  -1.69${}^{***}$ \\	   

\hline
\end{tabular}
\end{table}

\subsection{Veiling}

RU Lup exhibits both photometric and spectroscopic variations. An important
contributor to this variability is the veiling continuum. Veiling manifests
itself as a featureless continuum that decreases the depth of photospheric
absorption lines. The {\it veiling factor} is defined as the ratio of the
pseudo-continuum and the stellar continuum.

Part of the current dataset (the data from 2000) were used by Stempels \&
Piskunov (\cite{st02}) for a detailed analysis of the veiling in RU Lup. We
applied the same method on the entire dataset and found an average veiling
factor of 3.4, but with strong short-term variability. Some observations of RU
Lup obtained in 2005 have veiling factors as high as 9.

\subsection{Radial velocities}

We used traditional cross-correlation with a (synthetic) template spectrum over
the spectral region $6000$ to $6030$ {\AA}, which is entirely free from line
emission. Here, the S/N-ratio of our observed spectra is $\sim 125$, and
the $\chi^2$-statistics of our measurements yield typical errors of $\sigma =
\sqrt{2 / {\nabla^2 \chi^2}} \approx 0.16$~km~s${}^{-1}$. Correction for
possible instrumental drifts was done by cross-correlating the telluric oxygen
lines around $6300$ {\AA} with the corresponding lines in the NSO solar spectrum
(Kurucz et al. \cite{kurucz84}). The corrections turned out to be about $\pm
0.2$ km s${}^{-1}$ within a night. These corrections, as well as the
heliocentric velocity correction, were applied to each observed spectrum.
Finally, we obtained radial velocities for RU~Lup by cross-correlating the
(corrected) observed spectra with the synthetic template spectrum.

The obtained velocities, given in Table~\ref{tab:radvel}, range from about
$-4.0$ to $+1.0$ km s${}^{-1}$, which clearly shows that RU Lup displays
changes in radial velocity that are much larger than the velocity corrections
due to instrumental drifts or any other uncertainty.

We also checked the quality of our reduction method by performing a similar
analysis on spectra of a radial velocity standard star. The scatter in radial
velocity was less than 0.05 km s${}^{-1}$, indicating that the internal
precision of our radial velocity measurements is satisfactory.

\subsection{Line bisectors}
\label{sec:bisectors}

The observed amplitude of $v_{\rm rad}$ is $\sim$ 2.2 km s$^{-1}$. This is
well within RU~Lup's $v \sin i$ value of $9.0$ km s${}^{-1}$. It is thus
important to consider whether these are genuine line shifts or possibly due
to asymmetries within the line profile. 

At medium resolution or low $v \sin i$ it may be difficult to determine
asymmetries from individual line profiles. Instead of using single lines, an
``average'' line profile can be obtained from the cross-correlation function (CCF)
of the photospheric absorption lines. The bisector of the CCF then provides a
measure for the ``average'' line asymmetry. The method of calculating CCF
bisectors and its applications are described in more detail by Dall et al.
(\cite{dall06}). In our analysis we used the same CCF as the one we recovered
from the radial velocity measurements in the previous section, except that we
used a template without rotational broadening. Cross-correlation with such a
narrow template enhances any asymmetries in the CCF.

There exist several definitions of measures to quantify the asymmetry of a CCF
bisector. In our analysis we will use the so-called Bisector Inverse Slope (BIS,
see Queloz et al. \cite{queloz01}) defined as $\mbox{BIS} = v_t - v_b$, where
$v_t$ is the mean bisector velocity between 10\% and 40\% of the line depth, and
$v_b$ is the mean bisector velocity between 55\% and 90\% of the line depth. We
found that that there is a detectable line asymmetry in RU Lup, and that the BIS
varies between approximately $+1$ and $-1$~km~s${}^{-1}$, see Table
\ref{tab:radvel}. The relation between the BIS and the observed radial
velocities is discussed in Sect. \ref{sec:spots}.

\section{Periodicity}
\label{sec:periodsearch}

\begin{figure}[t]
\centering
\includegraphics[angle=90, width=\columnwidth]{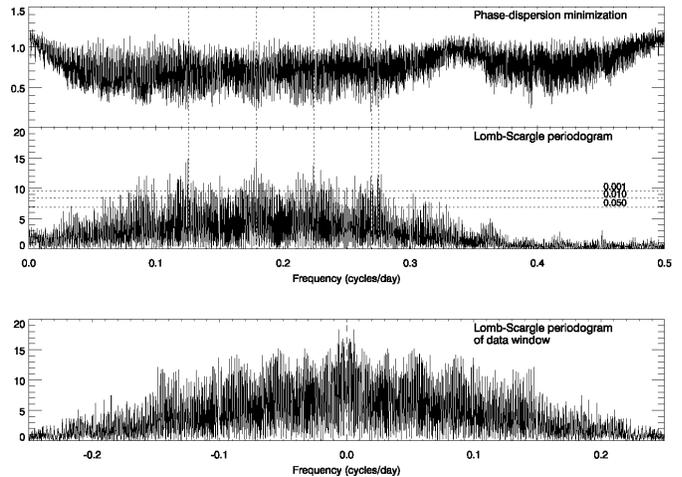}
   \caption{{\it Upper panel:} Comparison of the results of
   phase dispersion minimization and the Lomb-Scargle periodograms based on the
   observed data from Table~\ref{tab:radvel}. Some of the possible solutions,
   with periods of 3.63293, 3.71058, 4.45633, 5.58098, and 7.94913 days are
   highlighted. For the Lomb-Scargle periodogram, three levels of the
   false-alarm probability are indicated. {\it Lower panel:} The Lomb-Scargle
   periodogram of the data window function, with the peak at zero frequency
   removed. For purpose of clarity, a different frequency domain was used for
   this panel.
   }
\label{fig:periodograms}
\end{figure}

\subsection{A 3.7-day period in the radial velocity}

The relatively smooth changes in radial velocity within a night prompted us to
search for periodic radial velocity changes. We also searched the literature for
published radial velocity measurements of RU Lup, and we included data published
by Melo (\cite{me03}) in our analysis (see Table~\ref{tab:radvel}). Other
published velocities of RU Lup lack information on exact times of observation.

The combined dataset we have at our disposal is inhomogeneously sampled in time,
sometimes with several observations in short periods of time. The differences
between the radial velocities of consecutive spectra is small. Therefore, in
order to focus on the long-term evolution of the radial velocity, rather than to
contaminate the period analysis with the very short-term variations, we averaged
all consecutively observed observations, leaving us with a set of 38 radial
velocity measurements, including the 7 values published by Melo. For our
analysis we assumed an uncertainty in the radial velocities of $\sigma = \pm 1$
km s${}^{-1}$.

We analysed the data using two different period-detection methods. These methods
are the Lomb-Scargle periodogram (according to the prescription of Horne \&
Baliunas \cite{hb86}) and phase-dispersion minimization (Stellingwerf
\cite{pdm}). The resulting diagrams for these methods are shown in Fig.
\ref{fig:periodograms}.

An important feature of the Lomb-Scargle periodogram is that the height of the
peaks in the periodogram allows to make an estimate of the false-alarm
probability (FAP). We performed Monte-Carlo simulations to determine the
probability that, for a given signal in the periodogram, normally-distributed
random fluctuations give rise to a false detection of periodicity (i.e. the
FAP). In Fig.~\ref{fig:periodograms} we indicate the levels in the periodogram
corresponding to a FAP of 0.05, 0.01 and 0.001. The peaks in our periodogram
clearly exceed these levels, indicating that there is detectable periodicity in
our data, although it is not directly clear which period fits the data best.

\begin{figure}
\centering
\includegraphics[angle=90, width=\columnwidth]{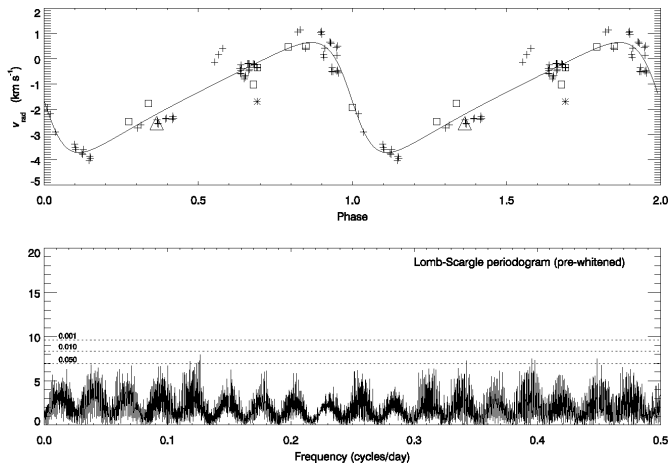}
   \caption{{\it Upper panel:} All observations of RU Lup, folded with the 3.71058 day
	    period. Observations marked with $+$ are taken at the VLT, with $*$
	    at the AAT and $\triangle$ at the CTIO. Observations by Melo
	    (\cite{me03}) are marked with $\square$. {\it Lower panel:} The
	    Lomb-Scargle periodogram after pre-whitening of the data with the
	    3.71058 day period. Levels of the false-alarm probability are
	    indicated.}
\label{fig:orbitwhitened}
\end{figure}

The difficulty in determining the correct period with the Lomb-Scargle
periodogram is for a large part due to the inhomogeneous distribution of our
observations in time, producing a broad window function with strong sidelobes.
(See Fig.~\ref{fig:periodograms}). Also, the Lomb-Scargle periodogram relies on
the implicit assumption that the signal is purely sinusoidal. Non-sinusoidal
signals might be detected, but with lower peaks of significance.
Phase-dispersion minimization is unbiased towards the shape of the radial
velocity curve, and might detect signals that are difficult to detect with the
Lomb-Scargle periodogram. 

As can be seen from Fig.~\ref{fig:periodograms}, the Lomb-Scargle periodogram
and Phase-dispersion minimization gave a number of possible solutions. Presented
with more than one clear solution,  we performed pre-whitening of the radial
velocity data and recalculated the periodograms. We found that the
pre-whitened periodogram of all periods contained significant levels of power,
except for the solution with $P = 3.71058$ days. The pre-whitened periodogram
for this period is shown in the lower panel of Fig.~\ref{fig:orbitwhitened}. The
remaining power is clearly less than the lowest levels of the FAP. After
pre-whitening, we can also estimate the uncertainty in the period (Horne \&
Baliunas, \cite{hb86}) finding that $P = 3.71058 \pm 0.0004$ days.

Although this period is well identified by  Phase-dispersion minimization, it
does not produce a strong peak in the Lomb-Scargle periodogram. However,
recalculating the Lomb-Scargle periodogram using {\it calculated} radial
velocities produced a diagram very similar to the periodogram of the observed
data, including peaks of other candidate periods found in the original
Lomb-Scargle periodogram This supports our conclusion that the most likely
period that fits our data has $P = 3.71058$ days.

If the period derived above is identical to the rotational period, this allows
us to determine the inclination of RU Lup. Herczeg et al. (\cite{herczeg05})
derived a radius of $1.64$ R${}_{\sun}$ and a mass of $0.65$ M${}_{\sun}$ from
the {\it stellar} luminosity, and if $P_{\rm rot} \approx 3.71$ days and $v \sin
i = 9.0 \pm 0.9$ km s${}^{-1}$ (Stempels \& Piskunov \cite{st02}) then $i
\approx 24\degr$. This is compatible with other observational arguments (Herczeg
et al. \cite{herczeg05}; Stempels \& Piskunov \cite{st02}) that indicate that RU
Lup is observed close to pole-on.

\subsection{Photometric periodicity?}

RU Lup is well-known to exhibit strong erratic photometric variability. Early
on, Hoffmeister (\cite{hoffmeister65}) and Plagemann (\cite{plagemann}) reported
a photometric period of 3.7 days. Giovannelli et al.
(\cite{giovannelli91}) found a weak signal corresponding to 3.7 days in the
power spectrum when re-analysing the early data, and Drissen et al.
(\cite{drissen89}) found some evidence of a 3.7 day period in their linear
polarization measurements. However, later analyses including newer data
indicated that the 3.7 day photometric period is not significant (Gahm et al.
\cite{gahm93}; Giovannelli et al. \cite{giovannelli94}).

From 2001 until 2005, RU Lup was also monitored in the $V$-band by the All Sky
Automated Survey (ASAS, see Pojmanski \cite{asas}). This period of time covers
most of our spectroscopic observations of RU Lup. We have analysed the ASAS
dataset of RU Lup, obtaining $V=11.23 \pm 0.26$, but no periodicity. Neither did
we find any short-term correlation between the measured radial velocities in
individual observing runs and simultaneous photometric data. We performed
several numerical tests on the ASAS data and found that the semi-amplitude of
any possible photometric periodicity must be less than 0.12 mag. Periodic
variations larger than this would have been detected from these data with high
probability.

\section{Discussion}

Periodic radial velocity changes in photospheric absorption lines may be
generated in different ways, and below we will discuss possible sources
such as rotational modulation of dark and/or bright spots on the stellar
surface, binarity and pulsations.

\subsection{Spots}
\label{sec:spots}

\begin{figure*}
\centering
\includegraphics[angle=90, width=\textwidth]{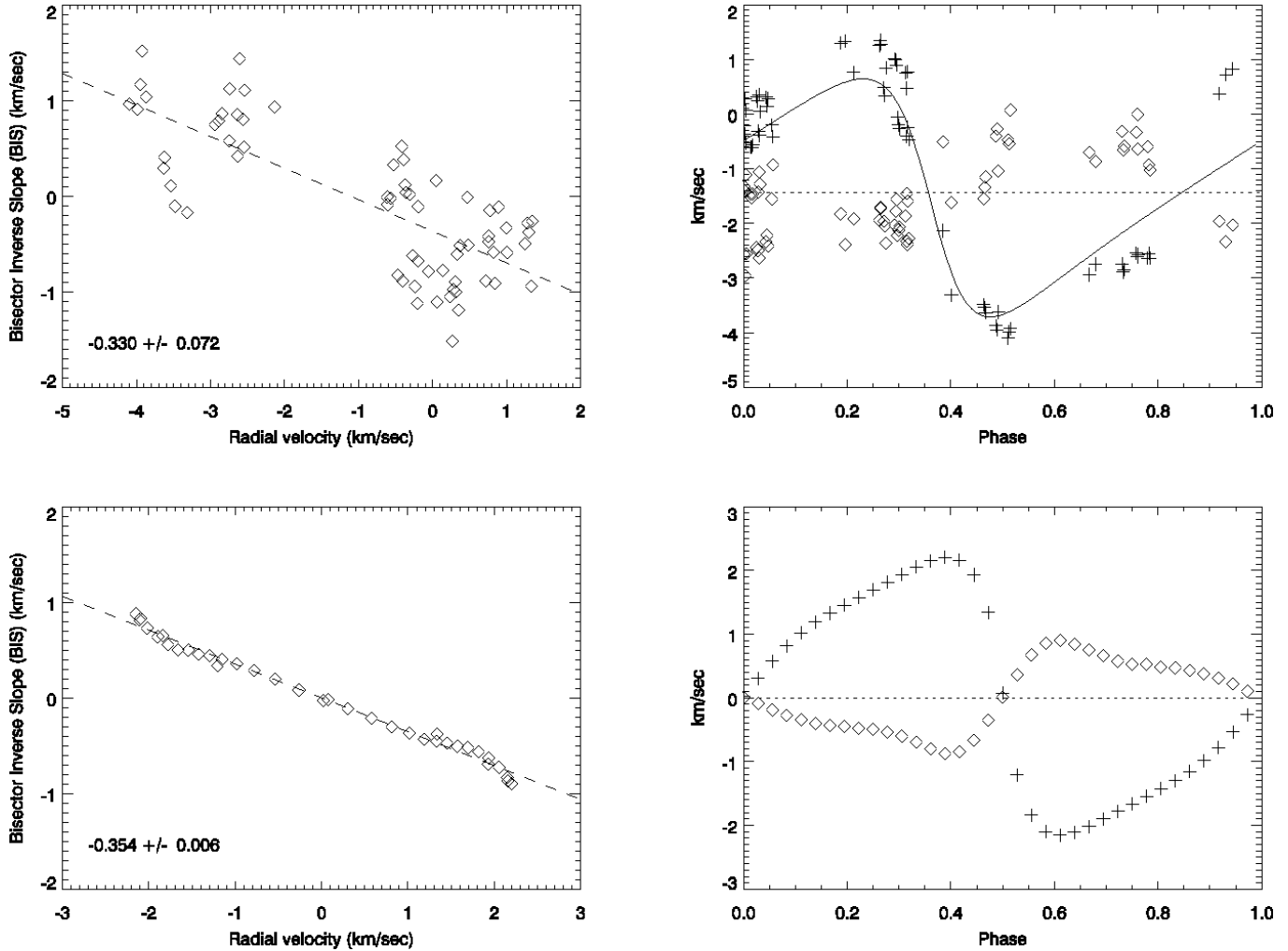}
\caption{
{\it Upper panel --- left:} The relation between the Bisector Inverse Slope
(BIS, see text) and the observed radial velocity of RU Lup. The slope of the
best least-squares fit is shown. {\it Upper panel -- right:} The observed radial
velocity ($+$) and the BIS ($\Diamond$) folded with the $3.71058$ day
period. The values of the BIS have been plotted with respect to the mean average
radial velocity. {\it Lower panel:} As in the upper panel, but for a set of
synthetic spectra of a star with a photosphere of $T = 4000$ K and a spot of $T =
3400$ K, a latitude of $60\degr$ and a radius of $35\degr$. 
}
\label{fig:bisectors}
\end{figure*}
 
Spots can distort the shape of photospheric line profiles, an effect
exploited in Doppler Imaging. Unfortunately, RU Lup has a $v \sin i$ of only 9
km s${}^{-1}$ making it unsuitable as a target for Doppler Imaging. Still, spots
may be an important candidate to consider as a cause of the observed velocity
changes in RU~Lup. 

A critical test of the effects of spots on the stellar line profiles can be
performed using line bisector analysis. An inhomogeneous distribution of spots on
the stellar surface will make the line profiles appear asymmetric, which is
detectable as a change in the line bisector. In addition to changes to the line
bisector, spots may cause variations in the apparent radial velocity of the
star, within a fraction of its $v \sin i$ value.

Any correlation between the BIS and the apparent radial velocity is often used
as a strong indication that the shape of the line profiles changes are due to
some process other than genuine shifts in radial velocity. For example, Queloz
et al. (\cite{queloz01}) used an anti-correlation between the BIS and the
observed radial velocity to show that spots can cause a periodic signal in the
apparent radial velocity. Similarly, Santos et al. (\cite{santos02}) used a
positive correlation between the BIS and the observed radial velocity to argue
that faint light from an undetected companion changes the intrinsic line
profile. Dall et al. (\cite{dall06}) also found a positive correlation for EK
Eri, and related this to surface activity, or possibly an unseen companion. The
absence of a correlation is normally interpreted as a strong argument for
genuine changes in radial velocity due to a companion (for example Santos et
al. \cite{santos03}).

In the top-left panel of Fig. \ref{fig:bisectors} we show the relation between
the BIS and the radial velocity we obtained for RU Lup. As one can see there is
an anti-correlation between the BIS and the apparent radial velocity of RU
Lup. The anti-correlation is more evident in the top-right panel, where we show
both the BIS and the radial velocity as a function of phase.

Although the anti-correlation between the BIS and the radial velocity is a
strong indication that surface structures might be responsible for the observed
changes in radial velocity, it is important to quantify this effect in terms of
spot size and changes in brightness. Therefore, we simulated the effect of
starspots on the shape of photospheric lines using a simple model of a spotted
star (see Hatzes \cite{hatzes02}).

\begin{figure}
\centering
\includegraphics[angle=0, width=\columnwidth]{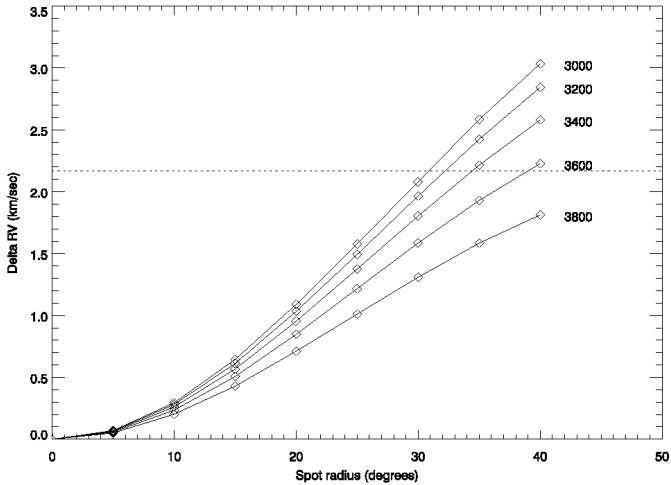}
\caption{
Calculated effect of a stellar spot located at a latitude of 60$\degr$ on the
amplitude of radial velocity variations for a range of temperatures and spot
radii. The dotted line indicates the observed amplitude of variations in RU Lup.
}
\label{fig:spots}
\end{figure}

In this model we assume parameters typical for RU Lup,  $T_{\rm eff}$ = 4000 K,
$v \sin i$ = 9 km s${}^{-1}$ and $i = 24\degr$. The photosphere was represented
by a synthetic K7 V spectrum, while for the spots we used synthetic spectra
across a range of temperatures (3000 -- 3800 K). The photospheric and spot
spectra were then combined to construct disk-integrated synthetic spectra of a
spotted star for several rotational phases and spot radii \mbox{(0 -- 40$\degr$)}. The
resulting spectra cover the spectral region 6000 -- 6050~{\AA} and have a
spectral resolution of about 60\,000. For each spectrum we determined the
corresponding radial velocity $v_{\rm rad}$, as well as the brightness and
the bisector shape through cross-correlation with the spectrum of an
unspotted K7 V star.

From our model we find that one single spot at a latitude of $\approx 60\degr$
and on an otherwise undisturbed stellar surface will cause the most dominating
effect (i.e. on radial velocity, brightness and line asymmetry). Although using
one single spot is the simplest possible model to reflect the observed
variations, this does not necessarily reflect the actual distribution. Doppler
Imaging of other CTTS has shown that a distribution with several (groups of)
spots is more realistic. We therefore interpret the results of our simulation
only as a first-order indication of the effect of an asymmetric distribution of
spots on the observed line profile.

The shift in observed radial velocity due to a spot is a function of both the
spot size and the temperature contrast between the spot and the stellar
photosphere, as is illustrated in Fig. \ref{fig:spots}. This figure shows that,
in order to explain the observed amplitude in radial velocity, the surface of RU
Lup must have a relatively large spot coverage (similar to a single spot with a
radius of $>30\degr$), and that these spots can have (average) surface
temperatures in the range 3000 -- 3600 K.

We did similar calculations for hot spots, assuming the spot to emit a
featureless spectrum (black-body continuum) of 7000~K (Calvet \& Gullbring
\cite{calvet98}). Because hot spots have a much larger surface brightness than
the rest of the photosphere, their contribution to the total spectrum is much
stronger than is the case for cool spots. We found that, although  asymmetric
hot spot may explain changes in radial velocity, the implied spot radius ($\ga
40\degr$) would cause very strong levels of veiling ($\ga 10$), which would
correlate strongly with phase. Hot spots would also cause strong phase-related
changes in the blue part of the spectrum. We found no evidence for either effect
in our observations of RU Lup. In addition, CTTS tend to have typical values of
hot spot filling factors around 0.1 -- 5\% (Calvet \& Gullbring \cite{calvet98};
Calvet et al. \cite{calvet04}), and not 25 -- 50\%. Non-polar hot spots also are
transient phenomena related to (variable) accretion. We find it therefore
unlikely that these short-lived, hot surface features could be responsible for
the long-term velocity variations in RU Lup. This, of course, does not rule out
active accretion and small hot spots on RU Lup.

Our simulations also allowed us to estimate the magnitude of brightness
variations due to spots. Although a cool spot with a radius of $40\degr$ will
induce a relatively strong modulation in brightness, it will be difficult to
detect this effect from photometry alone, since high levels of veiling will
strongly reduce the brightness contrast of such variations. This reduction in
brightness contrast between spot and star is also the reason why we were
unsuccessful in detecting other spectroscopic signatures of spots, such as
molecular bands from TiO.

For each set of simulated spectra we determined the behavior of the BIS, using
the same method and template as for the observed spectra. We found that the BIS
of the simulated spectra shows an anti-correlation with rotational phase, and
that the slope of this correlation depends weakly on both the spot size and
temperature. From our set of models, a star with a 35$\degr$ spot of $T = 3400$
K shows the best agreement with the observations, fitting the slope of the
anti-correlation, as well as the shape and amplitude of the velocity curve. This
is illustrated in the lower panels of Fig. \ref{fig:bisectors}.

We conclude that it is possible to explain the observed changes in radial
velocity in RU Lup with cool spots, and that this requires relatively large
surface areas covered with spots. The consistency of the radial velocity
variations, which allowed us to establish the rotational period, indicates a
long lifetime for the spot structures on RU Lup. Other CTTS for which
approximate spot sizes have been determined seem to follow this pattern. Doppler
Imaging studies of Sz 68 (Johns-Krull \& Hatzes \cite{johns97}), DF Tau (Unruh
et al. \cite{unruh98}), SU Aur (Petrov et al. \cite{petrov96}; Unruh et al.
\cite{unruh04} and MN Lup (Strassmeier et al. \cite{strassmeier05}) all recover
large surface areas covered by spots. Modelling of the spot coverage fraction in
active stars using molecular band analysis suggests that Doppler Imaging, and
thus also bisector analysis, may underestimate the true spot filling factor due
to features that do not change the line profile (O'Neal et al. \cite{oneil04}).

The large spot filling factor we find for RU Lup could also be in agreement with
the empirical relation between spot filling factor, photospheric temperatures
and magnetic field strengths on active dwarfs derived by Berdyugina
(\cite{berdyugina05}). This relation would imply a strong magnetic field of the
order of 3 kG in this star.

\subsection{Binarity}
 
One alternative explanation of the radial velocity variations in RU Lup may
be that a low-mass secondary component orbits the star.  An elliptic binary
orbit with $P = 3.71058$ days and $e = 0.41$ is compatible with the
periodic radial velocity variations observed in RU Lup (see
Fig.~\ref{fig:orbitwhitened}). Both the amplitude and the shape of the velocity
curve match the effect of a low-mass secondary orbiting RU Lup. 

The mass of the secondary can be determined under the assumption that the
axis of rotation are lined up, and that the system rotates synchronously. This
is a normal assumption for close binary systems, because the time-scale for
synchronization for such systems is relatively small (see Hilditch
\cite{hilditch} and references therein). The mass function for a single-lined
spectroscopic binary is $f(m) = {m_2^3} \sin^3 i / {(m_2 + m_1)^2} = 1.0361
\cdot 10^{-7}\,(1 - e^2)^{3/2} K_1^3 P$, where $e$ is the ellipticity, $K_1$ is
the amplitude of the velocity variations of the primary in km~s${}^{-1}$, and
$P$ the period of the system in days. Using $m_1 = 0.65$~M${}_{\sun}$, $K_1
= 2.17$ km s${}^{-1}$ and $i = 24\degr$, we derive a secondary mass of $m_2 =
0.027$ M${}_{\sun}$, which is well within the brown dwarf regime. The semi-major
axis of the primary follows from $a_1 \sin i = 1.9758 \cdot 10^{-2}\,(1 -
e^2)^{1/2} K_1 P$ giving $a_1 = 0.36$ R${}_{\sun}$ and $a_2 = m_1 / m_2 \cdot
a_1 = 8.5$~R${}_{\sun}$.
 
Although the orbital solution fits the observed radial velocity of RU Lup, we
find the solution of a brown dwarf secondary unlikely. The obtained solution
does not allow us to reproduce the observed correlation between the BIS and the
radial velocity. While a brown-dwarf binary seems unlikely, our observations
cannot dismiss a possible RS CVn-like scenario in which spots and a binary
companion cause simultaneous and phased modulation of the radial velocity.

\subsection{Pulsations}

The loci of fundamental pulsation periods of contracting stars over the
HR-diagram were calculated by Gahm \& Liseau (\cite{gahmlis}). The position of
RU Lup in the HR-diagram indicates possible periods on time-scales of hours
rather than days (Gahm \cite{gahm06}). Unless there is a not yet identified mode
of surface oscillations, it appears that pulsation is not the agent of periodic
radial velocity changes.

\section{Conclusions}

Using standard cross-correlation techniques we have found radial velocity
variations in very high resolution spectra of the CTTS RU~Lup. The variations
span a small range of velocities ($-4$ to $+0.5$ km s${}^{-1}$) and appear to be
periodic ($P \approx 3.71$~days). 
 
These radial velocity variations correlate with the slope of the bisector,
indicating that they are probably related to spots on the stellar surface. We
quantified this effect by simulating these variations with a model of a spotted
star. We find that a large part of the surface of RU Lup has to be covered with
long-lived spots of $T \approx 3400$ K to explain the observed changes in the
bisector and radial velocity. A large spot coverage indicates that a surface
magnetic field of about 3kG can be expected in RU Lup.

We also considered binarity and pulsations as other possible sources of radial
velocity variations, but these turned out to be unlikely candidates.

The 3.71 day period in radial velocity we recovered is similar to what was
reported in some very early photometric studies of RU Lup. Even though later
investigations, including this work, do not confirm any {\it photometric}
periodicity, it would be of value to repeat multi-colour monitoring, especially
in the blue part of the spectrum, to search for underlying weak
signals of periodic nature.

\begin{acknowledgements}

We thank John Barnes and Fred Walter for generously providing us their spectra
of RU Lupi. We are very grateful to the referee, Michael Sterzik, for the
constructive criticism that has helped us to interprete our data. This work was
supported by the Carl Trygger's Foundation. PPP acknowledges INTAS grant
03--51--6311.

\end{acknowledgements}

\end{document}